\documentclass[twocolumn,showpacs]{revtex4}

\usepackage{graphicx}%
\usepackage{dcolumn}
\usepackage{amsmath}
\makeatletter
\def\btt#1{\texttt{\@backslashchar#1}}%
\DeclareRobustCommand\bblash{\btt{\@backslashchar}}%
\makeatother

\begin{document}

\preprint{HEP/123-qed}

\title[Short Title]{An Improved Description of the Dielectric Breakdown in
Oxides Based on a Generalized Weibull distribution}

\author{U. M. S. Costa and V. N. Freire}                        %

\affiliation{Universidade Federal do Cear\'a, Departamento de F\'\i sica, Caixa
Postal 6030,\\
Campus do Pici, 60455-900 Fortaleza, Cear\'a, Brazil }

\author{L. C. Malacarne, R. S. Mendes and S. Picoli Jr.}

\affiliation{Universidade Estadual de Maring\'a, Departamento de F\'\i sica,\\
87020-900 Maring\'a, Paran\'a, Brazil}

\author{E. A. de Vasconcelos, E. F. da Silva Jr.}

\affiliation{Universidade Federal de Pernambuco, Departamento de F\'\i sica, \\
Cidade Universit\'aria, 50670-901 Recife, Pernambuco, Brazil.}

\date{\today}

\begin{abstract}

In this work, we address modal parameter fluctuations in
statistical distributions describing charge-to-breakdown
$(Q_{BD})$ and/or time-to-breakdown $(t_{BD})$ during the
dielectric breakdown regime of ultra-thin oxides, which are of
high interest for the advancement of electronic technology. We
reobtain a generalized Weibull distribution ($q$-Weibull), which
properly describes $(t_{BD})$ data when oxide thickness
fluctuations are present, in order to improve reliability
assessment of ultra-thin oxides by time-to-breakdown $(t_{BD})$
extrapolation and area scaling. The incorporation of fluctuations
allows  a physical interpretation of the $q$-Weibull distribution
in connection with the  Tsallis statistics. In support to our
results, we analyze $t_{BD}$ data of SiO$_2$-based MOS devices
obtained experimentally and theoretically through a percolation
model, demonstrating an advantageous description of the dielectric
breakdown by the $q$-Weibull distribution.

\end{abstract}


\maketitle



\section{Introduction}

The {\it International Technology Roadmap for Semiconductors}
indicates the need to decrease the SiO$_2$ gate thickness to less
than 3 nm to pursue the continuity of the Moore's Law \cite{ITRS}.
Research efforts to maintain this trend have been focusing on:
({\it i}) the advantageous substitution of SiO$_2$ by high
dielectric constant materials to allow  equivalent gate dielectric
oxide thickness bellow 1 nm \cite{high-k}; ({\it ii}) a better
understanding of the fractal regime of ultra-thin gate dielectric
oxide growth, where spatial inhomogeneities or fluctuations are
important \cite{fractal1,fractal2} considerations for the control
of dielectric breakdown conditions. Consequently, the reliability
of ultra-thin oxides for ultra-large scale integration is one of
the most important concerns in the domain of electronics
miniaturization nowadays \cite{Wu00,Degreave99, Green01}.

Two quantities which are experimentally measured to assess the
reliability of a metal-oxide-semiconductor (MOS) device are: the
charge-to-breakdown $Q_{BD}$, defined as the time-integrated
current density which flows until breakdown occurs, and the
time-to-breakdown $t_{BD}$, defined as the samples' lifetime.
These quantities are statistically distributed and are usually
assumed to be Weibull distributed \cite{Wolters86}. In particular,
accurate assessment of reliability of ultra-thin oxides is
seriously affected by fluctuations of oxide thickness. These
fluctuations cause a deviation from the Weibull distribution,
which is easily observed by a curvature in a graph of $\ln[-\ln
[1-F(t)]]$ versus $\ln[t]$, being\break $F(t)=\int_0^t P(x) dx$
the cumulative distribution. Extrapolation of the slope of this
plot from higher failure percentiles to lower failure percentiles
can lead to serious errors in reliability assessment, as shown by
Wu {\it et al.} \cite{Wu00}.

Deviations from a given distribution can be generated by taking
scale mixtures of it \cite{Gnedenko}. For instance, the scale
parameter of the exponential and Gaussian distributions can be
averaged by a gamma distribution. In these two cases, the
deviation induced by the fluctuation in this parameter has been
 successfully employed in the description of physical situations,
such as those related with nonexponential decay \cite{Wilk2,Wilk}
and  turbulence \cite{Beck,Batanov}. Furthermore, this average
process has also been connected with the entropic parameter $q$ in
the context of Tsallis statistics \cite{Beck,Wilk,Wilk2}. By
taking the fluctuations of the modal parameter of the Weibull
distribution into account, we obtain a generalized Weibull
distribution which properly describes $t_{BD}$ data when oxide
fluctuations are present, thus improving reliability assessment of
ultra-thin oxides by $t_{BD}$ extrapolation and area scaling. This
generalized Weibull distribution, without focusing fluctuations,
was discussed and applied in Ref. \cite{Sergio} to other systems.

In the following section we present the generalized Weibull
distribution employed here. The application of this distribution
to the description of the dielectric breakdown  in oxides is
considered in section 3.  The last section is dedicated to our
conclusions.

\begin{figure}
 \centering
 \DeclareGraphicsRule{ps}{eps}{}{*}
\includegraphics*[width=8cm,height=6cm,trim=1cm 0cm 0cm 0cm]{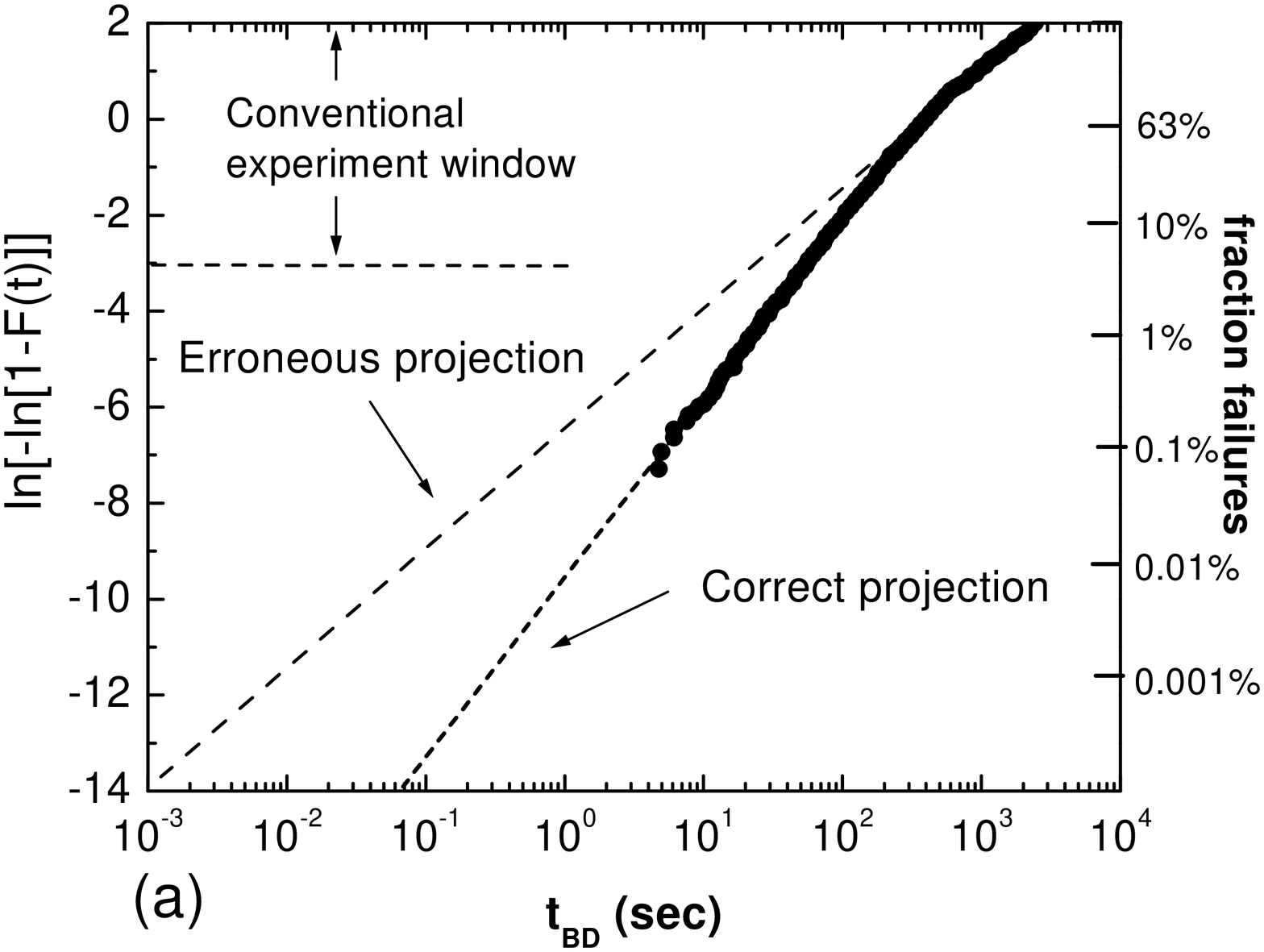}
\includegraphics*[width=8cm,height=6cm,trim=1cm 0cm 0cm 0cm]{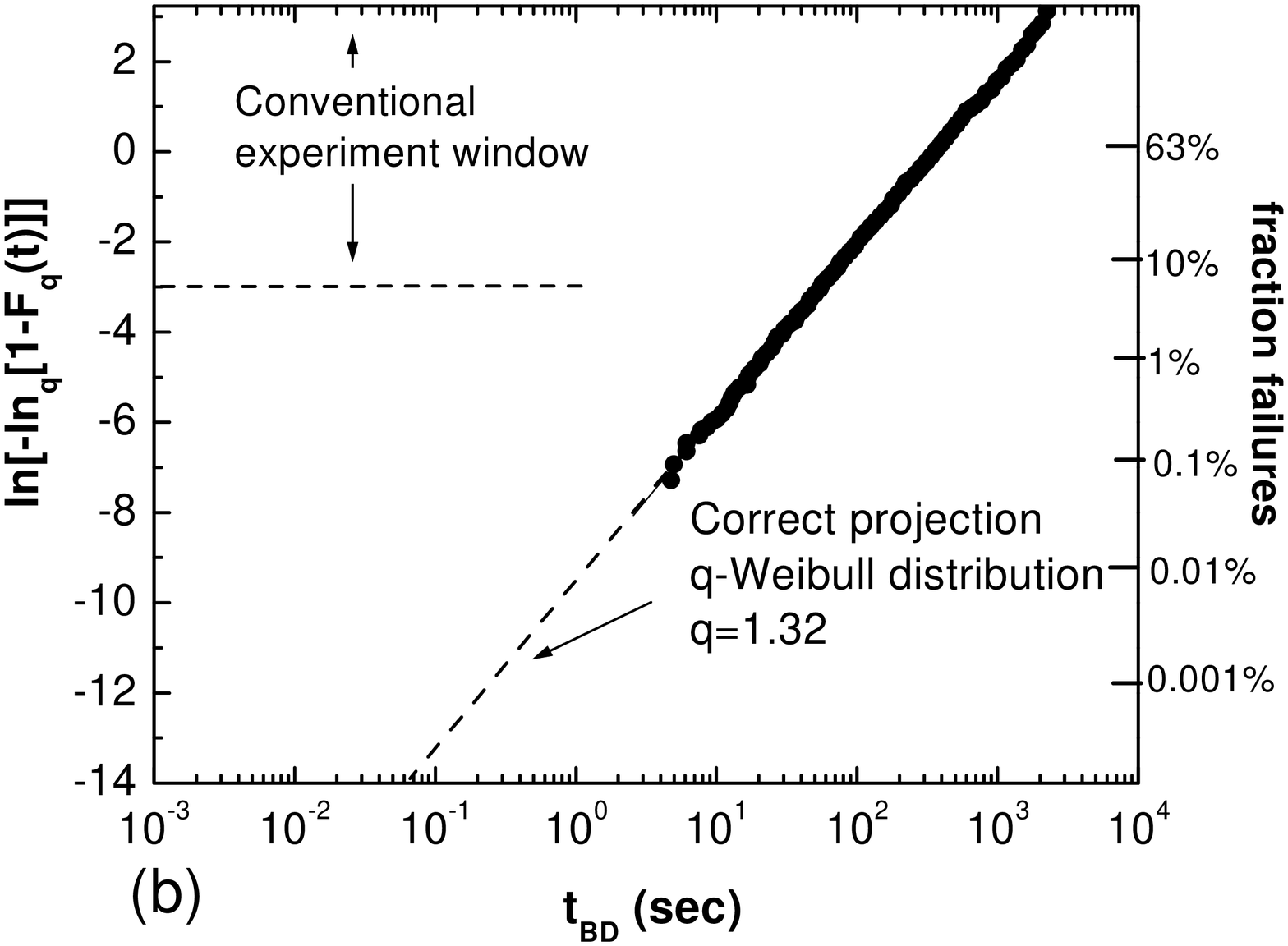}
 \caption{ (a) Graph of $\ln[-\ln[1-F(t)]]$ versus
$t_{BD}$ for 4000 ultra-thin oxide MOS capacitors. The data were
obtained from the work of Wu {\it et al.} \cite{Wu00}; (b) the
same data as in (a) but depicted in a $q$-Weibull plot in which a
correct $t_{BD}$ extrapolation is very clear.}
 \label{fig1}
\end{figure}

\begin{figure}
 \centering
 \DeclareGraphicsRule{ps}{eps}{}{*}
\includegraphics*[width=8cm,height=6cm,trim=1cm 0cm 0cm 0cm]{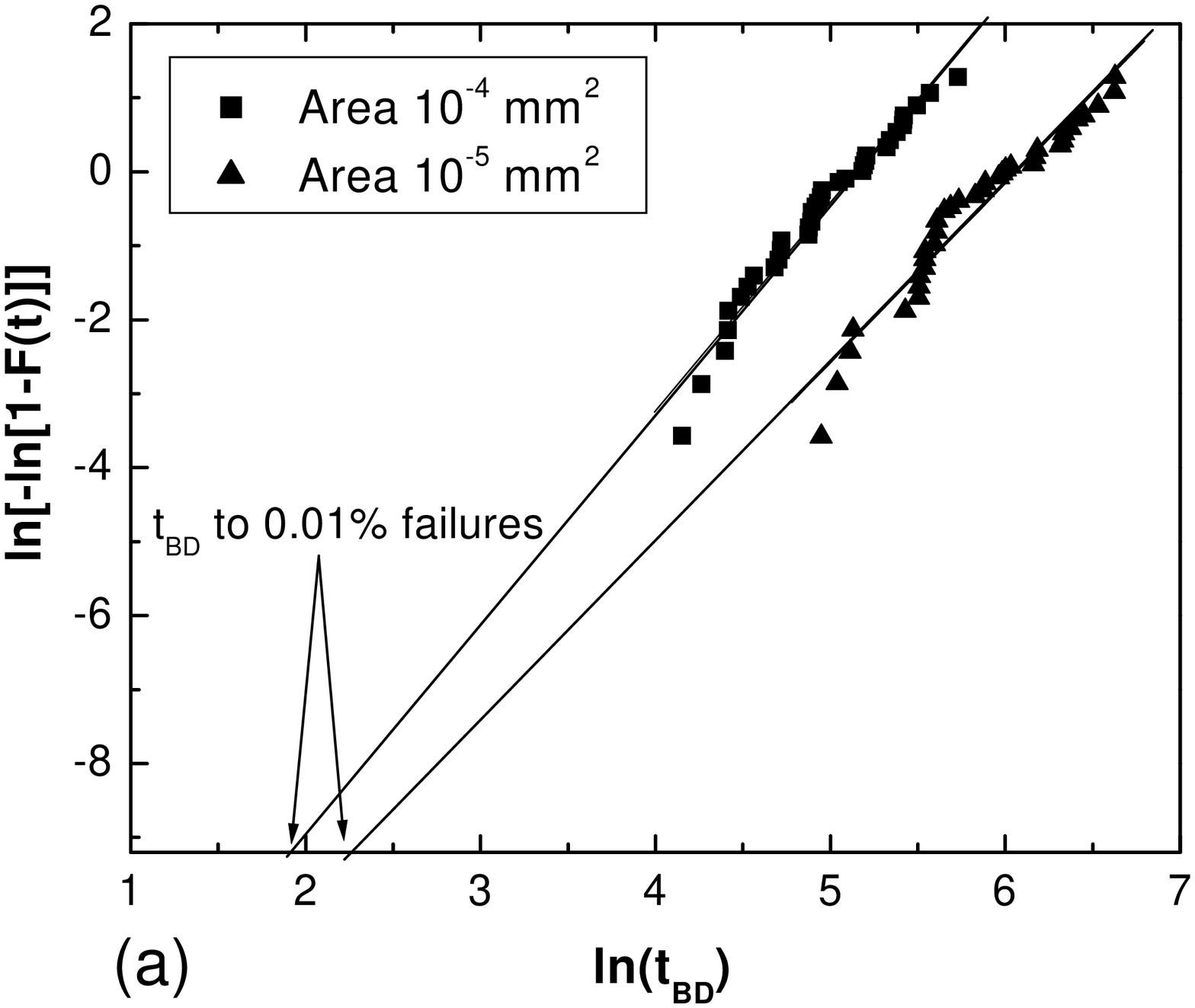}
\includegraphics*[width=8cm,height=6cm,trim=1cm 0cm 0cm 0cm]{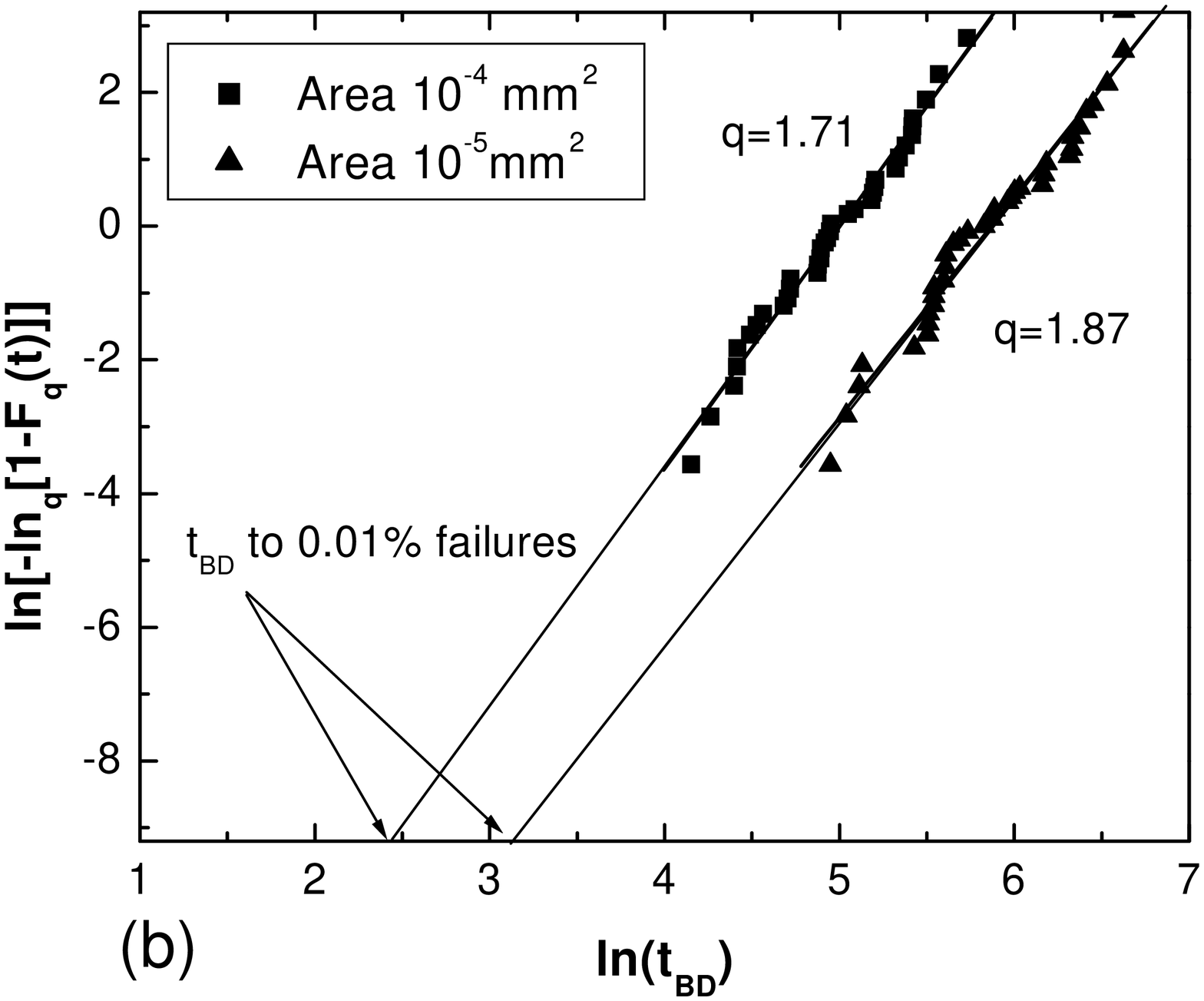}
\caption{(a) Graph of $\ln[-\ln_q[1-F^d(t)]]$ versus $\ln(t)$ for
a $6.9\thinspace nm$ sample oxides subjected to a
$E_{Ox}=12.7\thinspace MV/cm$ applied field. The samples have
areas of $10^{-4}$ and $10^{-5}mm^2$. The data were obtained from
the work of Teramoto {\it et al.}) \cite{teramoto}; (b) the same
data as in (a) but depicted in a $q$-Weibull plot showing improved
area scaling.} \label{fig2}
\end{figure}

\begin{figure}
 \centering
 \DeclareGraphicsRule{ps}{eps}{}{*}
\includegraphics*[width=8cm,height=6cm,trim=1cm 0cm 0cm 0cm]{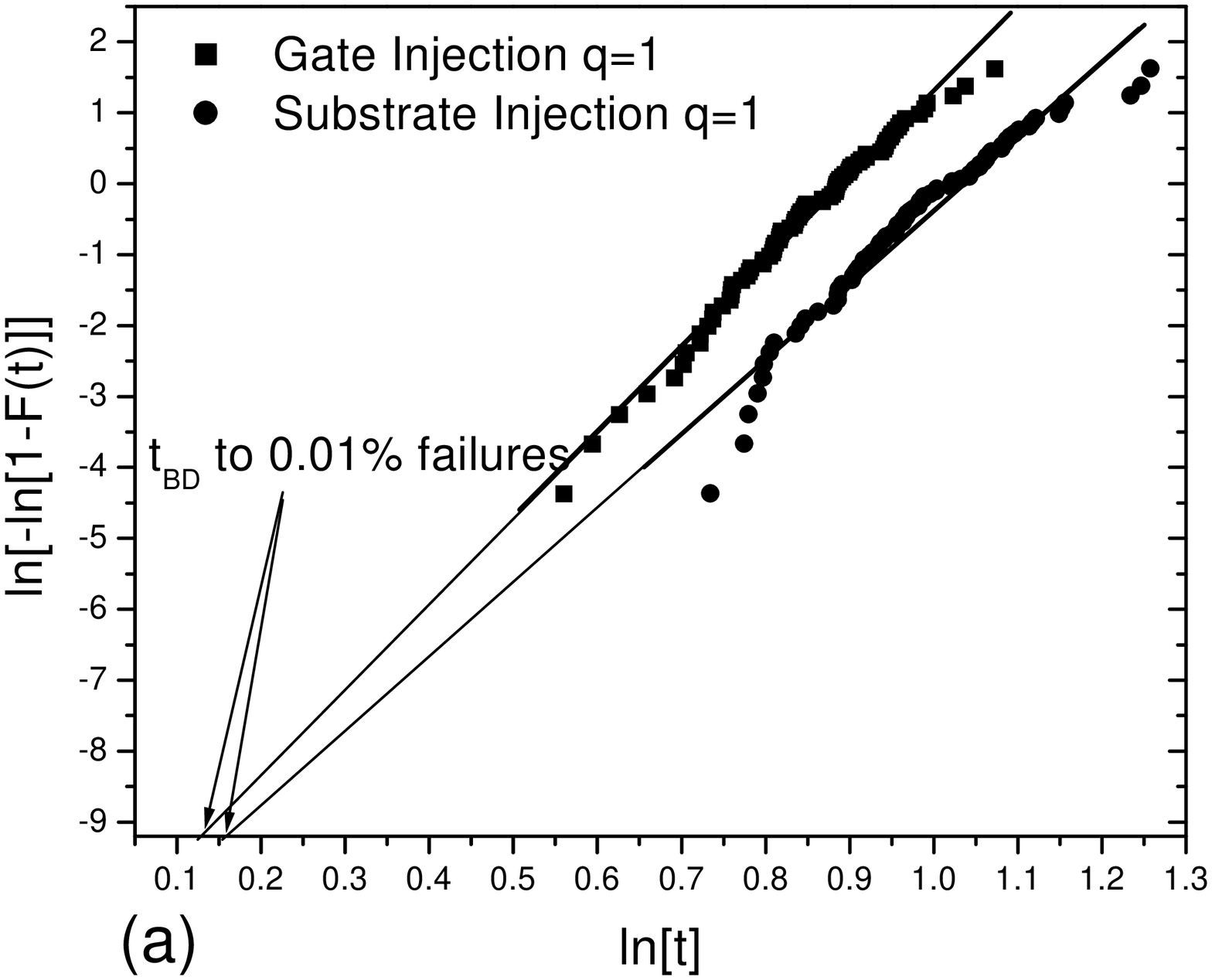}
\includegraphics*[width=8cm,height=6cm,trim=1cm 0cm 0cm 0cm]{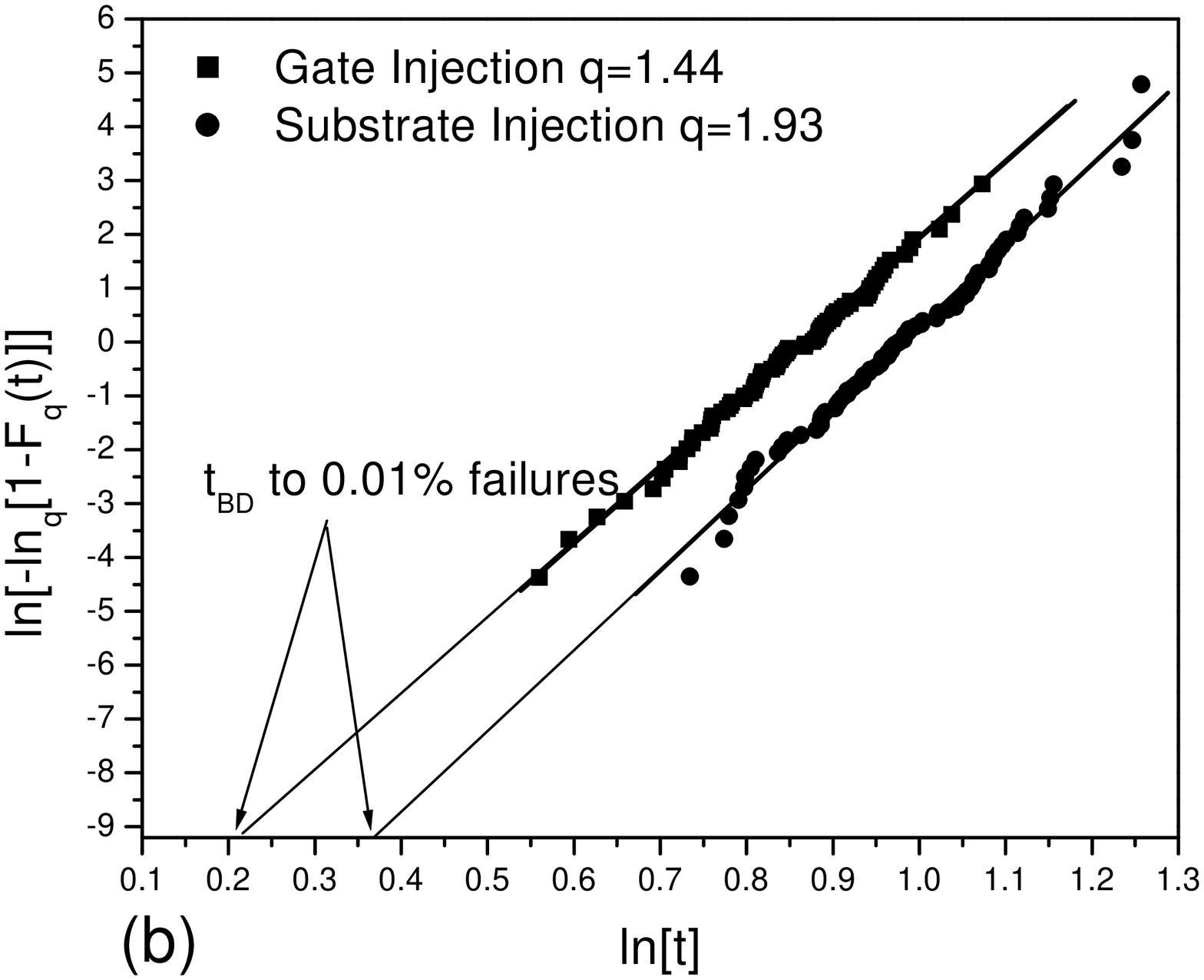}
 \caption{(a) Graph of $\ln[-\ln[1-F^d(t)]]$ versus $\ln(t)$ from
the percolation simulations as performed by Sombra {\it et al.}
\cite{sombra} for gate (squares) and substrate (dots) injection
adjusted with Weibull distributions; (b) the same data as in (a)
but depicted in a $q$-Weibull plot showing improved
fitting.}\label{fig3}
\end{figure}

\section{$q$-Weibull distribution}

The Weibull distribution often used in the statistical description
in the study of the time-to-breakdown in electronic devices
\cite{Degraeve00} is given by
\begin{equation}\label{Weibull}
P_1(x)=\frac{\beta}{x_0} \left(\frac{x}{x_0}\right)^{\beta-1}
\exp\left[-\left(\frac{x}{x_0}\right)^\beta\right],
\end{equation}
where the parameter $x_0$ is the modal value that corresponds to
$63,2 \%$ of the samples' lifetime. The parameter $\beta$ is the
Weibull slope and defines the time-to-breakdown spreading degree.
In order to incorporate the referred fluctuation, we consider a
Weibull compound distribution $p(x;\theta)= \beta \theta x^{\beta
-1} \exp (-\theta x^\beta)$. Here, $\theta=x_0^{-\beta}$, with
$\beta$ fixed and $x_0$ changing  so that the new parameter
$\theta$ is a stochastic variable ruled by the gamma probability
density function

\begin{equation}\label{beta}
  p(\theta)=\frac{\delta^r \theta^{r-1} e^{-\delta \theta}}{\Gamma (r)} \hskip
2cm (\theta, r,\delta > 0).
\end{equation}
Therefore, the average in the stochastic parameter $\theta$,
\begin{equation}\label{qw1}
\langle p(x;\theta) \rangle_\theta =\int_0^\infty p(x;\theta)
p(\theta) \;d \theta \; ,
\end{equation}
leads to \cite{Johnson}
\begin{equation}\label{qw}
\langle p(x;\theta) \rangle_\theta = r \delta^r \beta x^{\beta-1}
(x^\beta + \delta)^{-(r+1)}.
\end{equation}

Note that by naming this average  $P_{q'}(x)$, it can be written
as
\begin{equation}\label{q-W}
P_{q'}(x)= \frac{\beta(2-q')}{\tilde{x}_0}\left(
\frac{x}{\tilde{x}_0}\right)^{\beta-1} \exp_{q'}\left[ -
\left(\frac{x}{\tilde{x}_0}\right)^\beta\right],
\end{equation}
where we used the $q$-exponential function defined as
$\label{expq} \exp_{q'} (-a\, x) \equiv [1-(1-q')\,a
\,x]^{1/(1-q')}$ if $1-(1-q')\,a \,x\geq 0$, and $\exp_{q'} (-a
\,x)\equiv 0$ if $1-(1-q')\,a \, x<0$, with $q'=(2+r)/(1+r)$ and
$\tilde{x}_0^\beta =[ \delta/(r+1)]$.

By comparing Eq.(\ref{Weibull}) with Eq.(\ref{q-W}) we verify that
$P_{q'}(x)$ is a generalization of the Weibull distribution in the
same sense that the $q$-exponential function generalizes the
exponential one. In view of this observation and following Ref.
\cite{Sergio}, we refer to $P_{q'}(x)$ as $q$-Weibull
distribution. We also remark that $P_{q'}(x)$ contains a broad
class of distributions as particular cases. In fact, in the limit
$q' \rightarrow 1$ it reduces to the Weibull distribution, for
$\beta \rightarrow 1$  it gives the $q$-exponential distribution,
and when  $q' \rightarrow 1$ and $\beta \rightarrow 1$ it becomes
the exponential distribution. Furthermore, from Eq.(\ref{qw}) we
verify  two power law regimes, one for small $x$, $x^{\beta-1}$,
and another for large $x$, $x^{-(1+r\beta)}$. Notice also that $
\int_0^\infty P_{q'}(x)dx$ exists and is equal to one when $
q'<2$.

Before addressing our discussion to the application of the
$q$-Weibull distribution concerning experimental and simulation
results, we give a connection of this distribution with the
Tsallis statistics.  The first step towards verifying this
connection is based on the fact that the $q$-exponential can be
viewed as a signature of the  Tsallis statistics, since it
basically replaces the exponential in the canonical distribution.
In fact, this canonical distribution is obtained when the Tsallis
entropy $S_q \equiv (1-\sum_{i=1}^W p_i^q)/(1-q)$ is maximized
subjected to appropriate  constraints \cite{Tsallis}. Here, $p_i$
is the probability of the $i$th state, $W$ is the number of
accessible states,  and $q$ is a real parameter that rules the
degree of generalization of the theory (when $q\rightarrow 1$ we
recover the  usual entropy). On the other hand, the
$q$-exponential function can be obtained from the average of the
exponential function by using the gamma distribution, Eq.
(\ref{beta}) \cite{comment2}. Thus, the  parameter $q$ occurring
in the  Tsallis statistics is shown to be entirely induced by the
fluctuations of the parameter of the usual exponential
distribution \cite{Wilk2}. Following this interpretation, the
average process employed to obtain $P_{q'}(x)$ characterizes an
implicit connection with the  Tsallis statistics.

\section{Application to dielectric breakdown}

To apply our distribution, Eq. (\ref{q-W}), to investigate
experimental and numerical simulated data, we consider the
cumulative distribution for $P_{q'}(x)$, {\it i.e.},
\begin{equation}\label{q-rank}
F_q(t)=1-\exp_q\left[-\left(\frac{t}{\alpha}\right)^\beta \right],
\end{equation}
with $q=1/(2-q')$ and $ \alpha=\tilde{x}_0/(2-q')^{1/\beta}$. In
connection with the $q$-exponential function,  the $q$-logarithm
function is usually defined as \cite{Tsallis}
\begin{equation}
\ln_q(x)\equiv\frac{x^{1-q}-1}{1-q},
\end{equation}
thus $\ln_q[\exp_q(x)]=\exp_q[\ln_q(x)]=x$. By using Eq.
(\ref{q-rank}), we obtain that  the graph of
$\ln[-\ln_q(1-F_q(t))]$ versus $\ln(t)$ gives  a straight line
since
\begin{equation}
ln[-ln_q[1-F_q(t)]]\,=\beta\,ln(t)\,-\,\beta\ln(\alpha).
\end{equation}
Therefore, if a cumulative distribution relative to the dielectric
breakdown for a given set of data, $F^{d}(t)$, is well described
by the $q$-Weibull distribution, then  the $q$ parameter present
in the $q$-logarithm  is obtained through a linear adjustment, and
the parameter $\beta$ is the slope of the graph. This procedure
introduced here will be referred as a $q$-Weibull plot.

To compare the usefulness of the Weibull and the $q$-Weibull
distributions, we analyze experimental \cite{comment} and
simulation results. Figure \ref{fig1}(a), which is adapted from
the work of Wu et {\it al.} \cite{Wu00}, shows a curvature in a
graph $\ln[-\ln [1-F(t)]]$ versus $\ln[t]$. The variation of this
slope is associated with oxide thickness fluctuations. In fact, Wu
{\it et al.} \cite{Wu00} were able to generate this slope
variation by performing simulations in which the oxide thickness
obeyed a Gaussian distribution. Extrapolation of the slope of this
plot from higher failure percentiles (conventional experimental
window) to lower failure percentiles leads to an error in $t_{BD}$
projection and, consequently, in projection of reliability. It is
necessary to use large and time-consuming sample sizes (thousands
of devices) in order to obtain $t_{BD}$ distributions down to
lower percentiles and avoid erroneous projections. The same
$t_{BD}$ data can be correctly fitted in the entire range of
percentiles by the $q$-Weibull distribution, as shown in Fig.
\ref{fig1}(b). In the $q$-Weibull plot, the data follow a straight
line. We conclude that, by taking the fluctuations of the modal
parameter of the Weibull distribution into account, one obtains a
statistical distribution which completely describes $t_{BD}$ data
when oxide fluctuations are present. Therefore, the $q$-Weibull
distribution allows improved $t_{BD}$ projection for ultra-thin
oxides and does not require a very large number of samples, {\it
i.e.}, correct $t_{BD}$ extrapolation can be performed in the
conventional experimental window. This is a striking benefit
brought by the $q$-Weibull distribution.

In addition to correct $t_{BD}$ projection, we have found that the
$q$-Weibull also improves area scaling, another important
consideration for reliability studies \cite{Degraeve00}. In Fig.
\ref{fig2}(a), we have used  the data from the work of Teramoto
{\it et al.} \cite{teramoto}. They studied the time-dependent
dielectric breakdown for $6.9\thinspace nm$ thin SiO$_2$ oxides
with samples areas of $10^{-4}$ and $10^{-5}\thinspace mm^2$, and
subjected to a $E_{Ox}=12.7 \thinspace MV/cm$ applied field. It is
clear, when comparing  Fig. \ref{fig2}(a) and Fig. \ref{fig2}(b),
that the $q$-Weibull distribution gives a better area scaling than
the Weibull one, another remarkable benefit.

When $t_{BD}$ data are plotted in a limited range of percentiles,
the differences between the $q$-Weibull and the Weibull
distributions become less apparent, but  can still be seen. As an
example, we consider the simulations performed by Sombra {\it et
al.} \cite{sombra}. They  developed a percolation model to
describe the dielectric breakdown of a MOS capacitor investigating
effects of bias polarity, oxide film thickness and electric field
strength. The  hot electron injection is either through the gate
or the substrate. In Fig. \ref{fig3}(a) we have performed a linear
fit of the simulation data of Sombra \cite{sombra} with the
Weibull distribution, while in Fig. \ref{fig3}(b) we depict the
same data but with a linear fit through the $q$-Weibull
distribution. One can see once more that the $q$-Weibull
distribution gives a better fit than the Weibull one, for small
and large $t_{BD}$.

\section{Conclusions}

In summary, we conclude that:  ({\it i}) fluctuation on the modal
value is a mechanism to explain deviations from the Weibull
distribution in reliability studies of electronics devices; ({\it
ii}) this fluctuation leads to a generalization of the Weibull
distribution ($q$-Weibull distribution) and a connection with the
Tsallis statistics; and  ({\it iii}) the generalized Weibull
distribution leads to a better adjustment of experimental and
simulated data in comparison with the Weibull one, giving improved
$t_{BD}$ extrapolation and area scaling. Finally, the results
presented here can be very useful for an improved description of
the dielectric breakdown in the high dielectric constant materials
\cite{high-k} which will be present in future generations of
advanced MOS devices.


The authors would like to acknowledge the financial support
received from the Brazilian National Research Council (CNPq) under
contract CNPQ-NanoSemiMAt \#550.015/01-9, and the Ministry of
Planning (FINEP) through CTPETRO under contracts \# 65.00.02.80.00
and \#5000013/01-2, during the development of this work.




\begin{thebibliography}{10}
\bibitem{comment2} Strictly speaking, this average process only leads to
$q>1$,which is just the  relevant case to this work.
\bibitem{comment}The  experimental data were obtained through direct scanning of
the figures and then recovering the data using the Opti-Graph
program.
\end{thebibliography}
\end{document}